\documentclass[twocolumn,showpacs,preprintnumbers,amsmath,amssymb]{revtex4}

\usepackage{graphicx}
\usepackage{amssymb}
\usepackage{dcolumn}% Align table columns on decimal point

\begin{document}

\draft
\title{Correlation functions and queuing phenomena in growth processes with drift}
\author{S. Y. Yoon}
\author{Yup Kim} \email{ykim@khu.ac.kr  }
\affiliation{Department of Physics and
Research Institute for Basic Sciences, Kyung Hee University, Seoul
130-701, Korea}

%\date{\today}

\begin{abstract}
We suggest a novel stochastic discrete growth model which
describes the drifted Edward-Wilkinson (EW) equation $\partial h
/\partial t = \nu \partial_x^2 h - v\partial_x h +\eta(x,t)$. From
the stochastic model, the anomalous behavior of the drifted EW
equation with a defect is analyzed. To physically understand the
anomalous behavior the height-height correlation functions
$C(r)=\langle |h({x_0}+r)-h(x_0)|\rangle$ and $G(r)=\langle
|h({x_0}+r)-h(x_0)|^2\rangle$ are also investigated, where the
defect is located at $x_0$. The height-height correlation
functions follow the power law $C(r)\sim r^{\alpha'}$ and
$G(r)\sim r^{\alpha''}$ with $\alpha'=\alpha''=1/4$ around a
perfect defect at which no growth process is allowed.
$\alpha'=\alpha''=1/4$ is the same as the anomalous roughness
exponent $\alpha=1/4$. For the weak defect at which the growth
process is partially allowed, the normal EW behavior is recovered.
We also suggest a new type queuing process based on the asymmetry
$C(r) \neq C(-r)$ of the correlation function around the perfect
defect.

\end{abstract}

 \maketitle

%\begin{multicols}{2}

\section{Introduction}
Occasionally some of the research results on older topics give
excellent insights to the subsequent researches. One of the such
works is the recent work on the drifted Edward-Wilkinson (EW)
equation
\begin{equation}
\label{dEW} \frac{\partial h(x,t)}{\partial t} = \nu \partial_x^2
h (x,t) - v \partial_x h (x,t) + \eta(x,t)~,
\end{equation}
with a fixed boundary condition (FBC) $h(x_0 =0, t)=h(x_0 =L,
t)=0$ \cite{prue}. If the diffusion term $\nu\partial_x^2 h(x,t)$
is ignored in Eq. (\ref{dEW}), one can obtain roughness exponent
$\alpha=0$, growth exponent $\beta=0$ and dynamic exponent $z=1$
by a scaling argument and some other methods (\ref{dEW})
\cite{prue,sk1,sk2,sk3}. But for periodic boundary condition
(PBC), the drift term $-v\partial_x h(x,t)$ is irrelevant, and
thus Eq. (\ref{dEW}) becomes physically equivalent to the normal
EW equation \cite{ew}
\begin{equation}
\label{nEW} \frac{\partial h(x,t)}{\partial t} = \nu \partial_x^2
h (x,t) + \eta(x,t)~,
\end{equation}
with $\alpha=1/2$, $\beta=1/4$, and $z=2$. In contrast, if FBC
instead PBC is imposed, the roughness of surface described by Eq.
(\ref{dEW}) is exactly proved to show an anomalous behavior with
the exponents \cite{prue},
\begin{equation}
\label{aewx} \alpha=1/4,~~\beta=1/4,~~z=1 ~.
\end{equation}
The exact solution for Eq. (\ref{dEW}) \cite{prue} gives us an
excellent insight to understand the role of the drift term and the
boundary condition as a relevant extension of the normal EW
equation.

The drift term and the diffusion term are crucial for anomalous
exponents under FBC \cite{prue}. Without the drift term, the
maximum time which the roughness is fully developed for the system
is order of $L^2 /\nu$. If the system is affected by the drift,
the time scale that any noise-generated local structure starts
from one boundary to the other boundary is $t_{\times} =L/v$. The
fluctuation of the surface width is saturated after this time
scale. Therefore $W^2 \sim t_{\times}^{1/2} \sim L^{1/2}$ and thus
one can get the anomalous exponents (\ref{aewx}). In Ref.
\cite{prue}, local width is numerically studied to test the
validity of this physical mechanism. The local width is increased
as $W_l^2 (x_l)\propto {t_l}^{1/2}\sim ({x_l /v})^{1/2}$, with
$x_l$ being the position where the width is measured. But the
definition of the local width is not physically clear and there is
a sort of difficulty to measure $W_l^2 (x_l)$ for the small
system.

For last two decades, the stochastic discrete growth models whose
dynamical scaling behavior is identical to those of a given
continuum growth equation have been used to investigate the
dynamical scaling behavior for the corresponding surface roughness
\cite{ew,dyn,kpz,mh,le}. The stochastic discrete growth model is a
powerful tool which simplifies the complicate growth behavior and
provides an essential link between theory and experiments. For
example, the surface growth which is described by normal EW
equation has been studied through the Family model \cite{fam}. But
no simple growth model which exactly corresponds to the drifted EW
equation with FBC has been suggested yet. Other existing models
cannot explain the anomalous behavior of the surface fluctuation
which comes from the drift term and FBC in Eq. (\ref{dEW}). First
motivation of this paper is thus to make a simple stochastic
discrete model which satisfies Eq. (\ref{dEW}) with FBC using a
stochastic analysis method \cite{yoon}.

Second motivation of this paper is in the explanation of the
physical mechanism which causes the anomalous behavior
(\ref{aewx}) by using a height-height correlation function. The
height-height correlation function can be easily treated through
our handy stochastic discrete model. As mentioned above, the local
width \cite{prue} is somewhat unmature physical quantity and has
not been widely used for the analysis of surface roughness
\cite{dyn}. In contrast, height-height correlation function has
been extensively used to show the exact dynamic scaling in the
surface growth physics \cite{dyn}. The height-height correlation
function which we will use to analyze the surface with FBC is an
average of height differences or of squared height differences
between $h(x_0+r,t)$ and $h(x_0,t)$, where $r$ $(1\le r \le L)$ is
the distance from a boundary $x_0=0$ at time $t$. If we know the
height-height correlation function between two fixed boundaries,
we can get clean and clear information how the saturated surface
of whole system is formed. We will measure the height-height
correlation function and explain the anomalous behavior
(\ref{aewx}) through it.

Third motivation of this paper is to suggest about a new type
queuing phenomena originated from the analysis of the
height-height correlation function. Queuing \cite{dL} is a common
nonequilibrium phenomenon in nature. It is well established that
many driven flow processes belong to the same universality class
as Kardar-Parisi-Zhang (KPZ) type growth of one dimensional
interfaces \cite{kpz,rKPZ}. For example, a traffic jam \cite{Ha}
caused by slow and fast bond of the so-called asymmetric simple
exclusion process (ASEP) \cite{SN} is related to the faceting on
KPZ growth. ASEP breaks translational invariance in many ways. One
of them is to retain periodic boundary conditions, but to
introduce a defect into the system by modifying the transition
rates locally \cite{Ha,jn}. By properly chosen injection/removal
rates or defect strength, nonequilibrium phase transitions occur
between profiles of different shapes and average densities. In our
model, we map the fixed boundary condition to a defect which
exists on the center of system. The defect site means that the
growth at a certain site is impossible or is controlled by a
probability $p$ which is called a defect strength. That is, the
fixed boundary condition $h(x_0 =0,t)=h(x_0 =L,t)=0$ is shifted by
the mapping $x_0 \rightarrow x_0 +L/2$ to the defect site $h(x_0
=L/2,t)=0$ with the periodic boundary condition. For $p=0$ any
deposition or evaporation process at $x_0$ cannot occur
($h(x_0)=0$), since the strength of the defect is absolutely
strong and thus the surface heights cannot be freely increased by
the defect. If $p\neq0$, the deposition or evaporation process at
$x_0$ is accepted by $p$ and the system with $p=1$ is the same as
PBC without the defect. The height-height correlation function of
our stochastic discrete model shows lateral asymmetry $C(r)\neq
C(-r)$ around the defect, in contrast to KPZ-type queuing
phenomena or ASEP which shows up-down symmetry breaking under the
transformation $h(x) \rightarrow -h(x)$. The new queuing phenomena
from the drifted EW equation (\ref{dEW}) with a defect is
physically different from usual ASEP or queuing phenomena which is
related to the nonlinearity of KPZ equation \cite{kpz}, because it
is from the linearity of the drifted EW equation (\ref{dEW}).
Therefore it is very interesting to compare our new queuing
phenomena to earlier studied KPZ type queuing phenomena. In this
paper, we want to investigate new queuing phenomena from the
defect and the drift using our stochastic discrete model and apply
it to the parking garage model \cite{Ha}.

\section{A novel method to make stochastic discrete model}
In this section, we introduce a new method to make a discrete
stochastic growth model which physically corresponds to the
continuum equation. Normally the surface configuration is
described by the continuum Langevin equation \cite{dyn}
\begin{equation}
\label{gste} \frac{\partial h(x,t)}{\partial t}={\cal
F}(h(x),t)+\eta(x,t)~,
\end{equation}
where ${\cal F}(h(x) ,t)$ depends on the surface configuration
$\left\{ h(x)\right\}$. In the method, the surface configurations
are described in terms of integer height variables as $H\equiv
\{h_i \}_{i=1}^N ={0, \pm1, \pm2, \cdots}$ on a lattice. Then the
continuum equation (\ref{gste}) can be discretized into
\begin{equation}
\label{dste} \frac{\partial h_i}{\partial t}={{\cal F}_i}(h_i
,t)+{\eta_i}(t)~.
\end{equation}
Associated with Eq. (\ref{dste}), the general master equation
\cite{yup,risk,fox} for the surface evolution can be written as
\begin{equation}
\label{master} \frac{\partial P(H,t)}{\partial t} = \sum_{H'}
\omega (H',H)P(H',t)-\sum_{H'}\omega (H,H')P(H,t)~.
\end{equation}
Here $\omega (H',H)$ is the transition rate from the configuration
$H$ to the configuration $H'$ and $P(H,t)$ is the probability that
the system is in the configuration $H$ at time $t$. Using the
transition moments \cite{fox}
\begin{equation}
\label{tm1}K_i^{(1)} (H)=\sum_{H'} (h'_i - h_i )\omega (H' , H)
\end{equation}
and
\begin{equation}
\label{tm2} K_i^{(2)} (H)=\sum_{H'} (h'_i - h_i )(h'_j - h_j
)\omega (H' , H)~,
\end{equation}
the master equation (\ref{master}) can be rewritten as
Fokker-Planck equation \cite{yup,risk,fox}
\begin{eqnarray}
\label{fpe} \frac{\partial P(H,t)}{\partial t} &=& \sum_i
\frac{\partial}{\partial h_i} [K_i^{(1)} (H) P(H,t)] \cr &+&
\sum_{i,j} \frac{1}{2}\frac{\partial^2}{\partial h_i \partial h_j}
[K_{ij}^{(2)}(H) P(H,t)]~,
\end{eqnarray}
by Kramers-Moyal expansion. Therefore we can obtain the discrete
version of equivalent Langevin equation from Eq. (\ref{fpe}) as
\begin{equation}
\label{dLa} \frac{\partial h_i (t)}{\partial t} = K_i^{(1)} (H) +
\eta_i (t)~,
\end{equation}
and thus $K_i^{(1)}$ is equal to ${\cal F}_i$ of Eq. (\ref{dste}).
Here the noise $\eta_i (t)$ has the properties $<\eta_i (t)>=0$
and $<\eta_i (t)\eta_j (t')>=K_{ij}^{(2)}\delta(t-t')$. When the
noise is white, $K_{ij}^{(2)}$ is equal to $2D\delta_{ij}$.

Normally the evolution process in a stochastic surface growth
model is defined by one-particle deposition or evaporation in a
unit evolution process \cite{dyn}. Hence if only one-particle
deposition($h_i \rightarrow h_i +a$) or evaporation($h_i
\rightarrow h_i -a$) is imposed at a randomly chosen site $i$ in a
unit evolution process, then $K_i^{(1)}$ and $K_{ij}^{(2)}$ are
\begin{eqnarray}
\label{km} K_i^{(1)} (H)&=& a\omega_{id}-a\omega_{ie} \cr
K_{ij}^{(2)} (H)&=& \left\{ {0 \atop a^2 \omega_{id} + a^2 \omega
_{ie}} \quad{,i\neq j \atop ,i=j}\right.
\end{eqnarray}
from Eqs. (\ref{tm1}) and (\ref{tm2}), where $\omega_{id}$ is
$\omega(\{h_1, h_2, \cdots, h_i, \cdots, h_N\},\{h_1, h_2, \cdots,
h_i +a, \cdots, h_N\})$ and $\omega_{ie}$ is $\omega(\{h_1, h_2,
\cdots, h_i, \cdots, h_N\},\{h_1, h_2, \cdots, h_i -a, \cdots,
h_N\})$.Then from the Eq. (\ref{km}), these rates are
\begin{equation}
\label{tr1}
\omega_{id}=\left(\frac{K_i^{(1)}}{2a}+\frac{D}{a^2}\right)
\cdot\Theta\left(\frac{K_i^{(1)}}{2a}+\frac{D}{a^2}\right)~,
\end{equation}
\begin{equation}
\label{tr2}
\omega_{ie}=\left(-\frac{K_i^{(1)}}{2a}+\frac{D}{a^2}\right)
\cdot\Theta\left(-\frac{K_i^{(1)}}{2a}+\frac{D}{a^2}\right)~.
\end{equation}
Here we impose the unit step function $\Theta(x)$, i.e.
$\Theta(x)=1$ for $x> 0$ and $\Theta(x)=0$ for $x\leq 0$, to the
transition rates $\omega_{id}$ and $\omega_{ie}$, because the
transition rates must be always positive. If $\eta_i (t)$ is the
local white noise, the evolution site $i$ is randomly chosen to
apply the rates (12) and (13).

Now we want to describe the discrete stochastic model which
exactly correspond to Eq. (\ref{gste}) on an one-dimensional
($1d$) substrate in detail. Generalization to that on
higher-dimensional substrates can easily be obtained from the
description of the $1d$ model. The discrete model is defined by
the following steps. (i) Select a site $i$ randomly. (ii)
Calculate the deposition rate (the rate for $h_i \rightarrow h_i
+1$) $\omega_{id}$ and the evaporation rate (the rate for $h_i
\rightarrow h_i -1$) $\omega_{ie}$ from Eqs. (\ref{tr1}) and
(\ref{tr2}). (We set $a=1$.) (iii) Compare a random number
$R~(0<R<1)$ with the deposition probability
$P_{id}=\omega_{id}/(\omega_{id}+\omega_{ie})$ and the evaporation
probability $P_{ie}=\omega_{ie}/(\omega_{id}+\omega_{ie})$. If
$R\le P_{id}$, take the deposition process. Otherwise take the
evaporation process.

\section{Dynamical scaling properties of the stochastic model}

We now report the numerical results of our stochastic discrete
model for the drifted EW equation with the defect which is
controlled by a strength $p$. The initial surface is always set to
be flat ($h_i (t=0)=0$) and the lateral periodic condition is
always imposed. All data are taken by average over more than 500
independent runs. We first explain the scaling properties of the
surface width for several defect strengths $p$. The first
transition moment ${K_i}^{(1)}$ of drifted EW equation (\ref{dEW})
at a randomly selected site $i$ is
\begin{eqnarray}
\label{fmdE} {K_i}^{(1)} &=& [\nu_2 \nabla^2 h -v\nabla h]_i \cr
&=& \nu_2 [h_{i+1} + h_{i-1} -2h_i ]-\frac{v}{2} [h_{i+1}
-h_{i-1}].
\end{eqnarray}
We use the parameters $\nu_2 =1$, $v=1$ and $D=0.01$ for Eq.
(\ref{dEW}). We check the dynamical scaling behaviors for other
parameter values to obtain the same behaviors for $\nu_2 =1$,
$v=1$ and $D=0.01$. Used system sizes are $L=({2^6}+1) , \cdots,
(2^{12}+1) $ and the defect is located at the center site
${x_0}=(L-1)/2+1$ of all systems.

Usually the surface roughness is defined as $W
(L,t)=\sqrt{\langle\overline{h^2}\rangle -
\langle{\overline{h}}^2\rangle}$. The scaling relation of $W$
satisfies the following behavior \cite{dyn}
\begin{eqnarray}
\label{sc} W(L,t)&\sim& L^\alpha {\it f}(t/{L^{z}}) \cr &=&\left\{
{t^\beta~,\atop L^\alpha~,}\quad{t\ll L^{z} \atop t\gg
L^{z}~.}\right.
\end{eqnarray}
Figure 1(a) shows the obtained saturated surface width $W_{sat}
(=W(t \gg L^z))$ from the stochastic model for various system size
$L$ and defect strength $p$. $W_{sat}$ satisfies the scaling
relation $W_{sat}\sim L^\alpha $ with $\alpha=0.50(1)$ for $p=0.1$
and $p=1$. Since $p=1$ case is the same as that without the
defect, $\alpha$ should be EW value $\alpha=1/2$. Even though the
defect strength is sufficiently large ($p=0.1$), the surface
roughness still follows normal EW behavior. In contrast, for
perfect defect or $p=0$, the surface width $W$ follows the
anomalous behavior with $\alpha=0.25(1)$ instead of EW behavior.
The main plot of Fig. 1(b) shows that $W(L,t)$ for various $L$ and
$t$ collapses well to scaling function Eq. (\ref{sc}) for $p=0.1$.
We use the exponents $\alpha=0.50$ from the fitting of $W\sim
L^\alpha (t\gg L^z)$ in Fig. 1(a) and $\beta=0.25$ $(W\sim t^\beta
(t\ll L^z))$ as shown in the inset of Fig. 1(a). Therefore the
dynamic exponent is $z=\alpha/\beta=2$ which is the value of the
normal EW universality class. The results in Fig. 1(b) show that
the dynamical scaling property for $p\neq 0$ satisfies the normal
EW behavior. In contrast, the exponents are changed for $p=0$
(Fig. 1(c)). The time dependence of $W(L,t)$ of the model at $p=0$
is shown in the inset of Fig. 1(c). $\beta=0.249(1)$ in Fig. 1(c)
is obtained by applying the relation $W\sim t^\beta (t\ll L^z)$ to
$W(t)$ of the system with $L=2^{12}+1$. We also check the
dynamical scaling relation of Eq. (\ref{sc}) by plotting
$\ln(W/L^\alpha)$ against $\ln(t/L^z)$ in Fig. 1(c). The data
collapse well to the scaling function Eq. (\ref{sc}) with
$\alpha=0.25$ and $z=1$. The results in Fig. 1 clearly show that
the anomalous behavior occurs only when the perfect defect ($p=0$)
exists in the system. These results agree with the analytic result
of Ref. \cite{prue}. From these results, we confirm that a sudden
crossover from the anomalous behavior to the normal EW behavior
occurs as soon as $p$ deviates from $p=0$.

\begin{figure}
\includegraphics[scale=0.6]{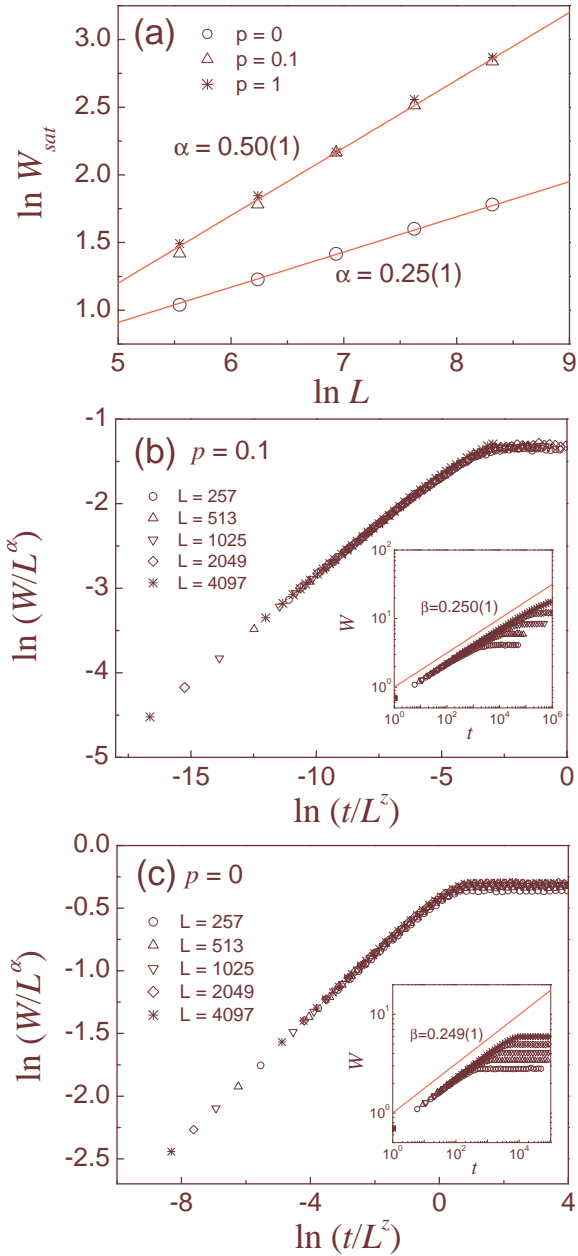}
\vspace{5mm} \caption{ (Color online) (a) Plot of $\ln W_{sat}$
against $\ln L$ in the saturation regime ($t\gg L^z$). The
straight lines represent the relation $W_{sat}\sim L^{\alpha}$.
$\alpha=0.25 (1), 0.5 (1), 0.5 (1)$ for $p=0$, $p=0.1$, and $p=1$,
respectively. Used system sizes are $L=(2^6+1) , \cdots,
(2^{12}+1)$ and used parameters are $\nu=1, v=1, D=0.01$. (b) The
plot of $\ln(W/L^\alpha)$ for the different $L$ against
$\ln(t/L^z)$ with $\alpha=1/2$ and $z=2$ at $p=0.1$. The numerical
data for $W$ nicely collapse to one curve which satisfies the
relation (\ref{sc}). (c) Scaling plot showing that the data for
$\ln(W/L^\alpha)$ plotted against $\ln(t/L^z)$ for various $L$
collapse to a single curve supporting the scaling function
(\ref{sc}) with $\alpha=1/4$ and $z=1$ at $p=0$. Insets of (b) and
(c) are log-log plots of $W$ against $t$. }
\end{figure}

\begin{figure}
\includegraphics[scale=0.5]{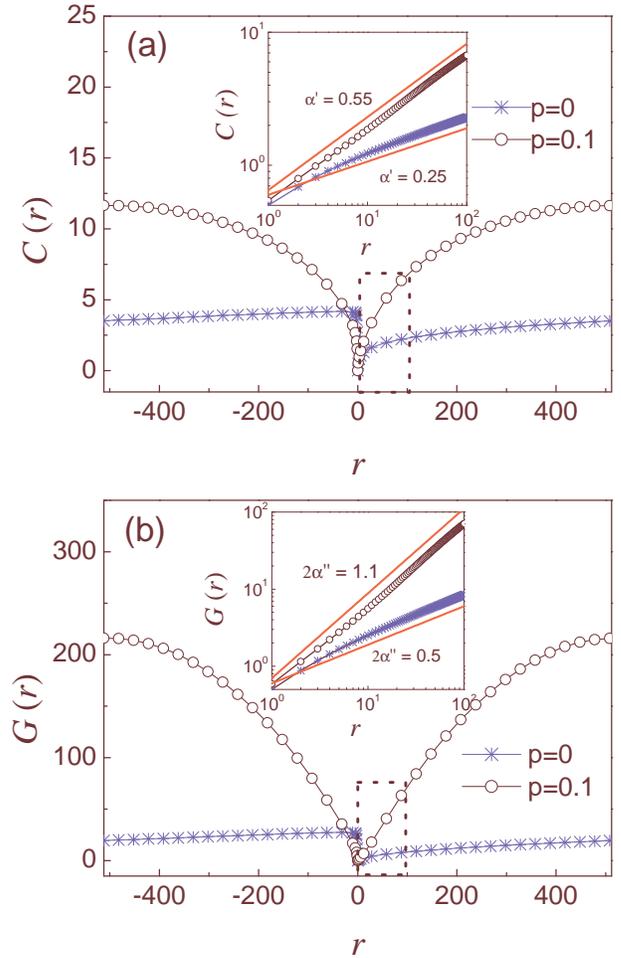}
\vspace{5mm} \caption{ (Color online) (a) Plots of the
height-height correlation function (\ref{corC}) for $p=0$ and
$p=0.1$ when $t\gg L^z$. Inset shows the log-log plot of the
dotted square part in the main plot. $C(r)$ satisfies the relation
$C\sim r^{\alpha'}$ with $\alpha'= 0.55(1)~(p=0.1)$ and $\alpha'=
0.25(1)~(p=0)$. (b) Plots of the height-height correlation
function Eq. (\ref{corG}) for $p=0$ and $p=0.1$. Inset shows the
log-log plot of the dotted square part in the main plot. $G(r)$
satisfies the relation $G\sim r^{2\alpha''}$ with $2\alpha''=
1.10(1)~(p=0.1)$ and $2\alpha''= 0.50(1)~(p=0)$. Used system size
is $L=1025$.}
\end{figure}

To understand the physical origin of the anomalous interface
profile which is induced from the defect (FBC) and the drift term
in Eq. (\ref{dEW}), we study several height-height correlation
functions. The height-height correlation function which has been
extensively used for the analysis of surface roughness \cite{dyn}
should be one of the efficient methods to see the physical
mechanism. Normally the height-height correlation function
\cite{dyn} means
\begin{equation}
\label{normalC} G_{PBC}(r,t)= \frac{1}{L}\sum_{x=1}^{L}\left<
[h(x,t)-h(x+r,t)]^2\right>~,
\end{equation}
when the periodic boundary condition is imposed. However when a
local defect is the relevant factor for the dynamical scaling
properties of the interface, the correlation function which
measures the height differences between $h(x_0)$ and
$h(x,t)$$(x\neq x_0)$ should be more powerful one. One of such
correlation functions is the naive and simple height-height
correlation function
$\widetilde{C}(r,t)=\left<h({x_0}+r,t)-h({x_0},t)\right>$. Here
$\left<~\right>$ means the average over independent runs and $r$
means the position from the defect $x_0(=L/2)$. However,
$\widetilde{C}(r,t)$ cannot characterize the surface configuration
physically well for the equilibrium surfaces in which the
deposition and evaporation processes occur with nearly the same
probability. The equilibrium surface fluctuates around $\left<h
\right>=0$ and $\widetilde{C}(r,t)$ has no information for the
surface height correlation around the defect. Instead, we use two
height-height correlation functions $C(r,t)$ and $G(r,t)$ defined
as
\begin{equation}
\label{corC} C(r,t)\equiv \left<|h({x_0}+r,t)-h({x_0},t)| \right>
\end{equation}
and
\begin{equation}
\label{corG} G(r,t)\equiv \left<|h({x_0}+r,t)-h({x_0},t)|^2
\right> ~,
\end{equation}
respectively. For the analysis of the saturated surface
configurations, we concentrate only on the saturated correlation
functions $C(r, t \gg L^z)(\equiv C(r))$ and $G(r,t \gg
L^z)(\equiv G(r))$. Used system size for the measurement of $C(r)$
and $G(r)$ is $L=1025$ and thus the defect site is located at
${x_0}=513$. These two height-height correlation functions are
expected to scale as $C(r)\sim r^{{\alpha}'}$ and $G(r)\sim
r^{2\alpha''}$ as shown in Figs. 2(a) and (b). Furthermore for
$p=0$ we obtain the exponents $\alpha'=\alpha''= 0.25(1)$, which
is very close to the roughness exponent $\alpha=0.25(1)$ in Fig.
1(a). For $p=0$,
\begin{equation}
\label{scC}C(r)=\left<|h(x_0+r,t)|\right>
\end{equation}
and
\begin{equation}
\label{scG} G(r)=\left<|h(x_0+r,t)|^2\right>
\end{equation}
and thus we clearly see the $\left<|h(x_0+r,t)|\right> \simeq
r^{1/4}$, which decide the anomalous behavior $\alpha =1/4$. This
means the saturated surface configurations for the drifted EW
equation with a defect satisfies
\begin{eqnarray}
\label{sW} W&\sim& \left< |h(x_0+r,t)|\right> \cr &\sim&
\sqrt{\left< h^2(x_0+r,t)\right> }~.
\end{eqnarray}
Therefore physically the surface configuration is decided by the
FBC (or the defect), which does not allow the surface fluctuation
freely. However, PBC allows the surface fluctuation freely even
though there exists the drift term $\partial_x h$.

In contrast, as soon as the growth at the defect is allowed even
with a very low probability, we get $\alpha'=\alpha''\approx 0.5$,
which is almost the same as roughness exponent $\alpha=0.5$ of EW
universality class. (See the data for $p=0.1$ in Fig. 2.)
Therefore the anomalous behavior $\alpha=\alpha'=\alpha''=1/4$ for
$p=0$ disappears and are changed into
$\alpha=\alpha'=\alpha''=1/2$ as soon as the defect condition is
released or for $p\neq 0$.

\begin{figure}
\includegraphics[scale=0.4]{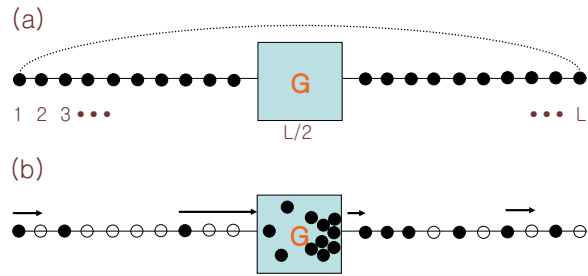}
\vspace{5mm} \caption{ (Color online) Schematic representation of
the queuing phenomena which emerges in the stochastic growth model
with a defect. Periodic boundary condition is used. (a) The
initial distribution of cars. (b) A steady state configuration.
Arrows denote the local velocities of cars. Cars move out from
garage with a very low speed and they are packed at the exit,
because of the defect. In the middle of the road, cars maintain
their average velocity. On the contrary, the speed with which cars
get in the garage is much faster than the average velocity.}
\end{figure}

\begin{figure}
\includegraphics[scale=0.5]{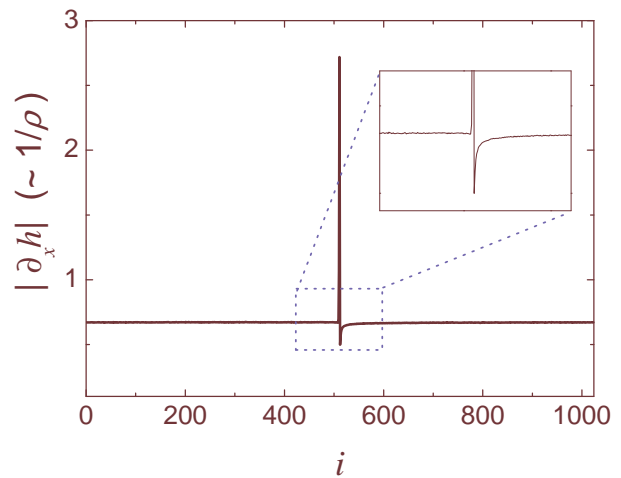}
\vspace{5mm} \caption{ (Color online) Plot of $|\partial_x h|$
against the position $i$. Used system size is $L=1025$. }
\end{figure}

Even though the defect exists, the drift term does not change the
universality of the dynamical surface scaling for $0<p<1$.
Therefore the height-height correlation function has a lateral
symmetry $C(r)=C(-r)$ or $G(r)=G(-r)$ for $p\neq 0$ as for the
normal EW equation (\ref{nEW}). For $p=0$, the noise-generated
surface structure from the defect is pushed by the drift and
transmitted all over the system. Then the flow of the surface
structure comes back through the boundary suddenly disappears at
the defect. Therefore the height-height correlation function shows
the asymmetry $C(r)\neq C(-r)$ and $G(r)\neq G(-r)$. This
asymmetric behavior of $C(r)$ and $G(r)$ is a unique behavior for
the drifted EW equation with a perfect defect. In contrast the
height-height correlation function around a defect for KPZ-type
surface growth \cite{Ha} always shows the lateral symmetry
$\widetilde{C}(r)=\widetilde{C}(-r)$. In KPZ-type surface growth
model with the defect \cite{Ha}, the height-height correlation
function $\widetilde{C}(r,p)$ changes from $\widetilde{C}(r,p)\sim
A|r|+B|r|^{\alpha'}$ for $p<p_c$ to $\widetilde{C}(r,p)\sim
B|r|^{\alpha'}$ for $p\ge p_c$. It means that the queueing effect
of the defect spreads out entire system ($p\ge p_c$) or affects
only around region of the defect ($p<p_c$). The phase transition
argument of KPZ-type surface growth is analogous to the queuing
effect by a defect which is located at a certain point of lane
\cite{Ha}. On the other hand, the height-height correlation
functions which we measured always maintains the simple power
behavior $C(r)\sim r^{\alpha'}$ and $G(r)\sim r^{\alpha''}$.
Instead, it shows the sudden crossover of the exponents $\alpha'$
and $\alpha''$ as soon as the defect is weakly imposed (or $p\neq
0$). That is, the effect of the defect and the drift influences
only when the defect strength is absolutely strong (or $p=0$).

\section{Application to queuing phenomena}
The asymmetric form of the interface profile by the defect and
drift can be applied to a kind of the queuing phenomena or car
condensation in a parking garage model \cite{Ha,jn}. Let's think
of the traffic flow of cars on an one-dimensional road with a
periodic boundary condition (or a ring-type road). The defect site
in our model is then reinterpreted as the parking garage (See Fig.
3). Initially all cars on the road are equally distributed
everywhere on the road as shown in Fig. 3(a). If the drift EW
equation (\ref{dEW}) with a defect (garage) describes a queuing
phenomena, the absolute gradient of the height $|\partial_x
h(t)|_i=|h_i(t)-h_{i-1}(t)|$ at $i$-th site can be mapped to the
effective speed $v_i$ of a car around the site $i$ at the given
moment $t$ due to the drift term $v|\partial_x h|$. In the steady
state the car density $\rho_i$ at the site $i$ is then inversely
proportional to $|\partial_x h|_i=|h_i-h_{i-1}|$ as $v_i \propto
|\partial_x h|_i \propto 1/\rho$.

Figure 4 shows $|\partial_x h|_i$ for $p=0$ in the saturated state
(or steady state) when $t \gg L^z$. $|\partial_x h|_i$ increases
very rapidly as $i$ increases from the defect position and
immediately reaches a certain average value. Average value remains
until $i$ approaches the defect site from the left, and
$|\partial_x h|_i$ grows very rapidly to the highest value in the
limit $i\rightarrow x_o-$. These results correspond to the
following steady-state traffic flow. The cars slowly move out from
the parking garage by a sort obstruction at the exit or by a lazy
parking charge collector. Thus they are packed near the exit of
the garage and the density $\rho$ of cars at the exit is very
high. As soon as cars enter the road, the speed and car density
reach to the average value $|\partial_x h|_i=v_a \sim
1/\rho_a\simeq const.$ and maintain the average values in the
middle of road. When cars come close to the entrance of the
garage, the speed of cars gets much faster. Therefore the density
of cars at the entrance becomes very low.

\section{Summary}
In this paper, we presented a simple stochastic discrete model
which describes the drifted EW equation (\ref{dEW}) with a defect.
From the simple model, we can easily show the anomalous behavior
of the surface fluctuation by the drift and defect. The scaling
exponents show the anomalous behavior $\alpha=1/4$, $\beta=1/4$,
and $z=1$, only when the defect strength is $p=0$. The exponents
are suddenly changed to normal EW exponents as soon as the defect
strength is weakly imposed ($p\neq 0$). We also measure the
height-height correlation functions to characterize the asymmetric
interface profile and to explain the exact physical role of the
drift term and defect. The height-height correlation functions
show that the drift and the perfect defect ($p=0$) makes the
asymmetry $C(r)\neq C(-r)$ or $G(r)\neq G(-r)$. These
height-height correlation functions satisfy the power law
$C(r)\sim r^{\alpha'}$ and $G(r)\sim r^{2\alpha''}$ with
$\alpha'=\alpha''=1/4$. The height-height correlation function
exponents and the roughness exponent have the same value
$\alpha=\alpha'=\alpha''=1/4$. Therefore the saturated surface
configuration for drifted EW equation with the defect is solely
decided by the defect. The asymmetry of the height-height
correlation functions is very unique when compared to the
correlation functions for KPZ-type growth model with the defect
which always shows the symmetry
$\widetilde{C}(r)=\widetilde{C}(-r)$. From this asymmetry we
suggest a new queuing process around the defect.

This work is supported by Korea Research Foundation Grant No.
KRF-2004-015-C00185.


\begin{references}

\bibitem{prue} G. Pruessner, Phys. Rev. Lett. {\bf 92}, 246101
        (2004).
\bibitem{sk1} Yup Kim and S. Y. Yoon, Phys. Rev. E {\bf 66},
031105 (2002).

\bibitem{sk2} Yup Kim and S. Y. Yoon, Phys. Rev. E {\bf 67},
056111 (2003).
\bibitem{sk3} Yup Kim and S. Y. Yoon, Phys. Rev. E {\bf 68},
036121 (2003).
\bibitem{ew} S. F. Edwards and D. R. Wilkinson, Proc. R. Soc.
        London Ser. A. {\bf 381}, 17 (1982).

\bibitem{dyn}
       {\em Dynamics of Fractal Surfaces},
        edited by F. Family  and  T. Vicsek
        (World Scientific, Singapore, 1991);
       A.-L. Barab\'{a}si and H. E. Stanley,
      {\em Fractal Concepts in Surface Growth} (Cambridge University
       Press, Cambridge, 1995); J. Krug, Adv. Phys. {\bf 46}, 139
       (1997);
       J. Krug and H. Spohn, in {\it Solids Far From Equilibrium:
       Growth, Morphology and Defects}, edited by C. Gordreche
       (Cambridge University Press, New York, 1991).
\bibitem{kpz} M. Kardar, G. Parisi and Y. -C. Zhang, Phys. Rev.
        Lett. {\bf 56}, 889 (1986).
\bibitem{mh} C. Herring, J. Appl. Phys. {\bf 21}, 301 (1950); W.
        W. Mullins, J. Appl. Phys. {\bf 28}, 333 (1957); {\bf 30}, 77
        (1959).
\bibitem{le} S. Majaniemi, T. Ala-Nissila and J. Krug, Phys. Rev.
        B {\bf 53}, 807 (1996).
\bibitem{fam} F. Family, J. Phys. A {\bf 19} L441 (1986).
\bibitem{yoon} S. Y. Yoon and Yup Kim, J. Korean Phys. Soc. {\bf
        44}, 538 (2004).
\bibitem{dL} B. Schmittmann and R. K.
        P. Zia, in {\em Phase transitions and
        Critical phenomena} vol. 17, edited by C. Domb and J. Lebowitz
        (Academic Press, London, 1995); G. M. Sch\"{u}tz, in {\em Phase transitions and
        Critical phenomena} vol. 19, edited by C. Domb and J. Lebowitz
        (Academic Press, London, 2001).
\bibitem{rKPZ} P. Meakin, P. Ramanlal, L. M. Sander and R. C. Ball, Phys.
        Rev. A {\bf 34}, 5091 (1986); M. Plischke, Z. Racz and D. Liu, Phys.
        Rev. B {\bf 35}, 3485 (1987); J. Neergaard and M. den Nijs, Phys. Rev.
        Lett {\bf 74}, 730 (1995).
\bibitem{Ha} M. Myllys, J. Maunuksela, J. Merikoski, J. Timonen,
        V. K. Horvath, M. Ha and M. den Nijs, Phys. Rev. E {\bf 68},
        051103 (2003).; M. Ha, J. Timonen and M. den Nijs, Phys. Rev. E
        {\bf 68}, 056122 (2003); B. Derrida, M. R. Evans, V. Hakim and
        V. Pasquier, J. Phys. A {\bf 26}, 1493 (1993)
\bibitem{SN} M. Schreckenberg, A. Schadschneider, K. Nagel and N.
        Ito, Phys. Rev. E {\bf 51}, 2939 (1995); K. Nagel, Phys.
        Rev. E {\bf 53}, 4655 (1996).
\bibitem{jn} S. A. Janowsky and J. L. Lebowitz, Phys. Rev. A {\bf 45}, 618
        (1992); B. Derrida, M. R. Evans and D. Mukamel, J. Phys. A {\bf
        26} 4911 (1993); B. Derrida, E. Domany and D. Mukamel, J. Stat. Phys.
        {\bf 69}, 667 (1992); S. A. Janowsky and J. L. Lebowitz, J. Stat. Phys.
        {\bf 77}, 35 (1994); F. J. Alexander, Z. Cheng, S. A. Janowsky and J. L.
        Lebowitz, J. Stat. Phys. {\bf 68}, 761 (1992); D. Kandel and D.
        Mukamel, Europhys. Lett. {\bf 20}, 325 (1992); G. Sch\"{u}tz, J. Stat. Phys. {\bf 71},
        471 (1993).

\bibitem{yup} Yup Kim, Prog. Theor. Phys. {\bf 104}, 495 (2000).
\bibitem{risk} H. Risken, {\em The Fokker-Planck Equation }
        (Srpinger-Verlag, Berlin, 1989).
\bibitem{fox} R. F. Fox and J. Keizer, Phys. Rev. A {bf 43}, 1709
        (1991).

\end{references}
\end{document}